\documentclass{article}
\usepackage{easy-todo}
\usepackage{url}
\usepackage{tikz}
\usetikzlibrary{arrows,automata,petri,shapes,calc,snakes}
\usepackage{amssymb,amsmath,amsthm}
\usepackage{xspace}
\usepackage{subcaption}
\newcommand{\step}{\mathtt{step}}
\newcommand{\obs}{\mathtt{obs}}
\newcommand{\dom}{\mathtt{dom}}

\newcommand{\purge}{\mathtt{purge}}

\newcommand{\out}{\mathit{out}}
\newcommand{\In}{\mathit{in}}

\newcommand{\snd}[1]{!_{#1}}
\newcommand{\rec}[2][]{?^{#1}_{#2}}

\newcommand{\machine}{\mathcal M}
\newcommand{\secpol}{\mathcal A}
\newcommand{\filters}{\mathcal F}

\tikzstyle{place}+=[inner sep=0,minimum size=3mm]
\tikzstyle{transition}+=[fill,minimum width=3mm,minimum height=0.5mm,font=\tiny]
\tikzstyle{synchron}=[fill,inner sep=0,minimum size=1.4mm,circle]
\tikzstyle{observ}=[draw,fill=white,line width=1pt,inner sep=0,minimum size=1.4mm,circle]
\tikzstyle{triggert}=[draw,fill,regular polygon,regular polygon sides=3,inner
     sep=0pt,minimum size=1.7mm,yshift=0.5mm]
\tikzstyle{triggerb}=[draw,fill,regular polygon,regular polygon sides=3,inner
     sep=0pt,minimum size=1.7mm,rotate=180,yshift=0.5mm]
\tikzstyle{triggerl}=[draw,fill,regular polygon,regular polygon sides=3,inner
     sep=0pt,minimum size=1.7mm,rotate=90,yshift=0.5mm]
\tikzstyle{triggerr}=[draw,fill,regular polygon,regular polygon sides=3,inner
     sep=0pt,minimum size=1.7mm,rotate=270,yshift=0.5mm]
\tikzstyle{component}=[fill=yellow!30,draw=orange,thick]
\tikzstyle{engine}=[fill=yellow!30,draw=orange,thick]
\tikzstyle{func}=[fill=gray!60,inner sep=0pt]
\tikzstyle{offer}=[thick,red]
\tikzstyle{notif}=[thick,black!70]
\tikzstyle{initial}=[after node path={(\tikzlastnode.225) edge[pre] ++(225:0.3)}]
\tikzstyle{initialUp}=[after node path={(\tikzlastnode.135) edge[pre] ++(135:0.3)}]
\tikzstyle{transitionh}=[transition,fill=black,minimum height=0.5mm,minimum width=3mm]
\tikzstyle{transitionv}=[transition,fill=black,minimum height=3mm,minimum width=0.5mm]

\title{Security policies for distributed systems\thanks{
   This work was carried out as part of the DMILS project \texttt{www.d-mils.org}  which is funded by the 
   European Commission under the contract Nr. FP7-ICT-3187727 of the $7^{\mathit{th}}$ Framework Programme for
   Information and Communications Technology. The smart grid case study has been supported by the Smart 
   Energy Systems action line of the EIT ICT Labs and the Bavarian Ministry of Economics.}}

\author{
Jean Quilbeuf
\and
Georgeta Igna
\and
Denis Bytschkow
\and     
Harald Ruess
}
\date{fortiss\\
              An-Institut Technische Universit{\"a}t M{\"u}nchen\\
              Guerickestr. 25, 80805 M{\"u}nchen, Germany \\
              \texttt{\{quilbeuf,igna,bytschkow,ruess\}@fortiss.org}}

\theoremstyle{plain}
\newtheorem{theorem}{Theorem}
\newtheorem{proposition}{Proposition}
\theoremstyle{definition}
\newtheorem{definition}{Definition}
\newtheorem{example}{Example}

\begin{document}

\maketitle

\begin{abstract}
A security policy specifies  a security property as the {\em maximal information flow}\@.
A {\em distributed system} composed of interacting processes implicitly defines an {\em intransitive} security policy 
by repudiating direct information flow between processes that do not exchange messages directly.  
We show that implicitly defined security policies in distributed systems are enforced, provided that processes 
run in separation,  and possible process communication on a technical platform is restricted to specified message paths 
of the system.
Furthermore, we  propose to further restrict the allowable information flow by adding {\em filter functions} for controlling 
which messages may be transmitted between processes, and we prove that locally checking filter functions is sufficient
for ensuring  global security policies.
Altogether,  global intransitive security policies are established by means of local verification conditions for the (trusted) processes 
of the distributed system.
Moreover,  security policies may be implemented securely on distributed integration platforms which ensure partitioning.  
We illustrate our results with a {\em smart grid case} study, where we use CTL model checking for discharging local verification 
conditions for each process under consideration.
\end{abstract}

\section{Introduction}

Modern applications are often implemented in terms of distributed systems consisting of interacting components.
Depending on the requirements of the application, some components  possess/know sensitive data or assets that must 
not be disclosed to a certain subset of the other components.
Consider, for example,  a voting system with a central unit for counting votes and voters.  The central unit needs to get hold of any
particular vote for counting, but it must not leak particular votes to other voters.

Such a {\em security property} may be expressed in terms of a {\em security policy}, which decomposes a system into
various security domains (e.g. users or processes) with consistent levels of information~\cite{gm82}\@. 
A security policy defines a maximal bound on the allowable information being transmitted between different security domains.
In this way, a security policy specifies how information is allowed to flow between different security domains, and,
for every possible run of the system,  security domains must not gather more information than what is allowed by the security policy. 

More concretely, we are considering distributed systems consisting of processes communicating through asynchronous message passing.
Such a distributed system implicitly determines a security policy by considering the processes of this system as security domains. 
Whenever two processes exchange messages, this implicit security policy allows information to flow from the sender of the message to its receivers. 
This implicitly defined security policy is  {\em intransitive}, since information may not necessarily flow directly from process $\pi_1$ to process $\pi_3$
whenever it flows directly from $\pi_1$ to $\pi_2$ and from $\pi_2$ to $\pi_3$\@. 
In these cases, the transmission of information from $\pi_1$ to $\pi_3$  requires transmission via $\pi_2$\@.



Intransitive security policies have previously been studied by Rushby~\cite{rushby1992} and van der Meyden~\cite{meyden07}\@. 
One of the main underlying assumptions of their framework is that any action can always be taken in any global state. 
Consider, for example, an action $a$  that transmits information from $\pi_1$ to $\pi_2$, an action $b$ that transmits
 information from $\pi_2$ to $\pi_3$, and  assume that $b$ forwards the information about $\pi_1$ that $\pi_2$ has gathered. 
Action $b$ can be executed before or after $\pi_1$ executed the action $a$, which yields different states in $\pi_3$, depending 
on the state of $\pi_2$\@. 
In order to capture this situation,  Rushby~\cite{rushby1992} and van der Meyden~\cite{meyden07} define the information available 
to a security domain in a recursive manner.
In our example, the information available in $\pi_3$ after executing $b$ contains the information available in $\pi_2$ before executing $b$\@.

In contrast to  Rushby~\cite{rushby1992} and van der Meyden~\cite{meyden07} we are working in an asynchronous setting with actions
for emitting and receiving messages. 
A system simply defines a set of possible  execution traces (e.g. Mantel~\cite{mantel00}, Balliu~\cite{balliu13})\@. 
Furthermore, we assume that messages contain all the information, that is, the information 
detained by a process is the complete history of received and sent messages. 
Therefore,  in contrast to Rushby~\cite{rushby1992} and van der Meyden~\cite{meyden07} , sending the message $b$ from 
the domain $\pi_2$ is possible only if $\pi_2$  has already received message $a$\@. 
When $\pi_3$ receives $b$, it may infer, knowing the set of possible executions, that $a$ has been transmitted.
We consider such as system to be secure, as the information was relayed by $\pi_2$\@. 
Such a system is considered to be insecure, however, whenever $\pi_3$ may observe directly the emission of $a$\@. 

The notion of implicit security policy for distributed systems as outlined above needs to be strengthened.
Consider, for  example,  a voting system, where each voter receives the result of the vote from the central unit once the election is done.
Since each voter sent his vote before the results are published, the implicit security policy allows the resulting 
message to contain the detail of the votes.
Therefore, the implicit security policy does not ensure that no voter may know the vote of another voter.
Chong and van der Meyden~\cite{chongmeyden12} introduce {\em filter functions} to strengthen a given security policy.
A filter function restricts the information between two security domains, depending to the history of the actions executed.
For instance, the filter function between the central unit and a voter might allow transmission of information only if
(1) all voters have voted and (2) the data sent back are only the election results.
The filter function of Chong and van der Meyden~\cite{chongmeyden12} replace actions,
whereas Zhang~\cite{zhang11} considers Boolean filter functions for allowing or disallowing information to flow.

We consider Boolean filter functions as in Zhang~\cite{zhang11}, and we require that a filter function depends only on the local history 
of the domain that can issue the information.
For instance, for the voting system example, the filter function depends only on the messages received and sent by the central unit.
A filter function restricts the outputs of one particular process.
Checking that the behavior of this particular process respects the filter functions, that is does not send messages it is not allowed to send, can be done separately.
We show that it is sufficient to check that every filter function is respected to prove that the security policy is met.
Components whose output is not restricted by a filter function do not need to be checked.
In the voting system example, we do not need to check the behavior of each voter, it is sufficient to check the central unit.

Our approach is illustrated by means of prosumer-based smart micro grid.
The smart micro grid negotiates with a set of prosumers --- that is, agents which may either produce or consume energy ---  in 
order to  assure the stability of the smart grid. 
We encode this prosumer-based smart grid as a distributed system and the main security property, that no
prosumer should be able to deduce the consumption or production of any other prosumer, 
is encoded by means of an intransitive security policy using Boolean filter functions.
We use model checking techniques on the trusted security domain to prove that the smart 
micro grid respects the specified filter function.

Security policies may be enforced by means of  time, space, and I/O partitioning. 
Separation kernels~\cite{rushby81} enable processes to share ressources (e.g. computing, memory) without introducing unwanted and 
hidden communication channels.
In a similar way,  time-triggered networks~\cite{kopetz05} provide time partitioning and ensure isolation of communication channels. 
A distributed, mixed-criticality platform combining these two approaches is currently being developed in the D-MILS project with the 
explicit goal of implementing and enforcing security policies on a distributed platform. 
In this way, security policies can be implemented on a distributed computing platform.

This paper is organized as follows. 
In Section~\ref{sec:framework}, we define our model of distributed systems, that we call \emph{distributed machine} and the notion 
of security policy. 
In Section~\ref{sec:unwinding}, we show that a distributed system complies with its implicit security policy and
that local check of filter function is sufficient to ensure that the global system meets a filtered security policy.
Our smart micro grid case study is detailed in Section~\ref{sec:case_study}\@. 
Finally, we present related work in Section~\ref{sec:related} and conclude with Section~\ref{sec:conclusion}\@.

\section{Information Flow}
\label{sec:framework}

The information flow of a system is the information exchanged between different security domains of a system.
A security policy defines a maximal information flow.
A system complies with  a security policy if no part of the system can obtain more information than allowed by the maximal information flow.

Following the work by Chong and van der Meyden~\cite{chongmeyden12}, we model a system using the notion of machine, and the maximum information flow as a security policy that are formally defined in the next sections. 
We formally state the fact that a machine complies with a security policy.

\subsection{Distributed Machine}

In this paper, we focus on distributed systems made of independent processes exchanging messages.
Our definition of distributed machine has been inspired by the machine model introduced by Rushby~\cite{rushby1992} and van der Meyden~\cite{meyden07}. 
We start by formally defining processes and then explain how to compose them.

\begin{definition} A process is a tuple
  $\pi = (S, s^0, A^{\In}, A^{\out}, \step)$, 
  where
  \begin{itemize}
 \item $S$ is a set of  states, with $s^0$ as the initial state, 
 \item $A = A^{\In} \cup A^{\out}$ is the set of actions composed of \emph{receive actions} $A^{\In}$  and \emph{send actions} $A^{\out}$, and, 
 \item $\step : S \times A \rightarrow S$ is a partial transition function. 
   We write $\step(s,a) = \mathit{undef}$ when the action $a$ cannot be executed from state $s$.
   \end{itemize}
\end{definition}
We distinguish between two types of actions in a process.
A send action, denoted $\snd{m}$ models the emission of the message $m$. 
A receive action, denoted $\rec[\pi]{m}$ models the reception of the message $m$ in the process $\pi$.
The set of states and the set of actions may be infinite.

Not all actions are possible from every state. 
We denote by $LocalExec(\pi)$ the sequence of actions that are valid executions of $\pi$.
For a sequence $\delta \in LocalExec(\pi)$, we denote by $s^0.\delta$ the state reached after executing the sequence $\delta$.
We denote by $\epsilon$ the empty sequence.

\begin{figure}[htbp]
  \centering
  \tikzstyle{lbl}=[font=\small,inner sep=2pt]
\tikzstyle{place}+=[minimum size=20pt]
  \begin{tikzpicture}
    \def\larg{1.9}
    \node[place,initialUp] (m0) {$h$};
    \node[place] (m1) at ($(m0) - (\larg,0)$) {$hr$} 
      edge[pre,bend left=10]   node[lbl,above] {$?^S_\text{res}$} (m0)
      edge[post,bend right=10] node[lbl,below] {$!_\text{display}$} (m0) ;
    \node[place] (m2) at ($(m0) + (\larg,0)$) {$hc$} 
      edge[pre,bend left=10]   node[lbl,below] {$?^S_\text{cmd}$} (m0) 
      edge[post,bend right=10] node[lbl,above] {$!_\text{cmdH}$} (m0);
    \node[place] (m3) at  ($(m0) - (0,1.9)$) {$\ell$} 
      edge[pre,bend right=20] node[lbl,right] {$?^S_\text{toggle}$} (m0) 
      edge[post,bend left=20] node[lbl,left] {$?^S_\text{toggle}$} (m0);
    \node[place] (m4) at ($(m3) - (\larg,0)$) {$\ell r$} 
      edge[pre,bend left=10]   node[lbl,above] {$?^S_\text{res}$} (m3)
      edge[post,bend right=10] node[lbl,below] {$!_\text{display}$} (m3) ;
    \node[place] (m5) at ($(m3) + (\larg,0)$) {$\ell c$} 
      edge[pre,bend left=10]   node[lbl,below] {$?^S_\text{cmd}$} (m3) 
      edge[post,bend right=10] node[lbl,above] {$!_\text{cmdL}$} (m3);
  \end{tikzpicture}
  \caption{Graphical representation of a process.}
  \label{fig:process}
\end{figure}
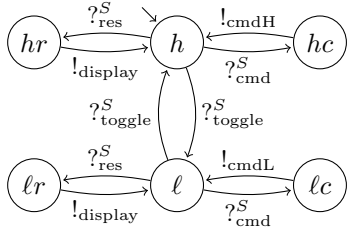


\begin{example}
  In Figure~\ref{fig:process}, we represent a process, that operates as a switch for sending commands either to a high or a low security network.
  This example shows a simplified version of the switch component from the Starlight Interactive Link~\cite{anderson96}.
  In~\cite{chongmeyden12}, the Starlight system is taken as an example to illustrate filter functions.
  This process may receive 3 types of messages, namely res, cmd and toggle.
  The reception of a toggle message (from the user) switches the security mode between high (state $h$) and low (state $l$).
  Whenever a cmd (command) message is received (from the user), it is forwarded to a network for execution.
  If the process is in high security mode, the command is sent to the high security network through a cmdH message, otherwise it is sent to the low security network through a cmdL message.
  Whenever a res (result) message is received, it is forwarded to the user through a display message.

  Formally, the set of states is $\{h,hr,hc,\ell,\ell r,\ell c\}$.
  The set of input actions is $\{\rec[S]{\text{cmd}},\rec[S]{\text{toggle}},\rec[S]{\text{res}}\}$ and the set of output actions is $\{\snd{\text{cmdL}},\snd{\text{cmdH}},\snd{\text{display}}\}$.
  The $\step$ function is defined as shown on the figure, for instance $\step(h,\rec[S]{\text{toggle}}) = \ell$ and $\step(hr,\rec[S]{\text{toggle}}) = \mathit{undef}$.
\end{example}


In this paper, we consider \emph{distributed machines} that are defined as the composition of a set of processes.
The processes communicate through asynchronous message passing. 
An action of the distributed machine is either the emission or the reception of a message.
As the synchronous reception of the message might not be possible in every state of a process, we assume that each process is equipped with a buffer for storing incoming messages.
The consumption of a message in the buffer corresponds to a receive action.
Identifying the sender of an action allows mapping this action to a security domain which is needed to reason about information flow.
Therefore, we require that, for each message defined in the set of processes, there is a unique sender.
\begin{definition}[Process composability]
  Given a set of processes $\{\pi_1,\ldots,\pi_n\}$, where for each $i$, $\pi_i = (S_i,s^0_i,A^{\In}_i,A^{\out}_i,\step_i)$, we say that they are \emph{composable} iff
  for each message $m$, there exists a unique process $\pi_i$ such that $\snd{m} \in A^\out_i$ and at least one process $\pi_j$ such that $\rec[\pi_j]{a} \in A^\In_j$. 
  We call process $\pi_i$ the \emph{sender} of $m$ and $\pi_j$ a \emph{receiver} of $m$.
\end{definition}

The behavior of a distributed machine is as follows.
Sending a message $\snd{m}$ is done by executing the transition labeled by $\snd{m}$ in the sender and adding $m$ in the buffer of each receiver of $m$.
Receiving a message $\rec[\pi_i]{m}$ in $\pi_i$ is done by removing an occurrence of $m$ from the buffer of $\pi_i$ and executing the corresponding transition locally.
The buffer is represented by a sequence of input actions.
If several receive actions are possible, only the one corresponding to the message occurring first in the buffer can be executed.

A machine as defined in~\cite{meyden07} is a transition system, extended with a decomposition into security domains.
Each of these security domains delimits a subpart of the system that is granted a given level of information.
The definition of a machine contains an observation function stating what each domain can observe from the global state.
In a distributed machine, each process corresponds to a security domain, which can only observe its local state and the buffer of incoming messages.

\begin{definition}
  Given a set of processes $\{\pi_1,\ldots,\pi_n\}$ that are composable, a \emph{distributed machine} is a tuple $\machine  = (S,s^0,A,\step,D,\dom,\obs)$, 
  such that 
  \begin{itemize}
    \item $S = (S_1 \times  {A_1^{\In}}^*) \times \ldots \times (S_n \times {A_n^\In}^*)$, and $s^0=((s^0_1,\epsilon), \ldots, (s^0_n,\epsilon))$. 
      We denote by $(q_1,\ldots,q_n)$ a global state of the system, where for each $i$, $q_i = (s_i,\beta_i)$ indicates the current state $s_i$ of the process $\pi_i$ and the contents $\beta_i$ of the input buffer.
    \item $A = \bigcup\limits_{i=1}^n \left( A^\In_i \cup A^\out_i\right)$ 
      The actions of the distributed machine are emissions and receptions of messages.
 \item $\step$ contains two types of transitions, that correspond to sending and receiving messages:
   \begin{itemize}
     \item if for a process $\pi_i$ there exists $\snd{m} \in A_i^{\out}$ such that $\step_i(s_i,\snd{m}) \neq \mathit{undef} $ then  
       $\step( (q_1,\ldots,q_n)$, $\snd{m}) = ( q'_1, \ldots , q'_n)$ where
       \[ q'_j = (s'_j,\beta'_j) =
	 \begin{cases}
	   (\step_j(s_j,\snd{m}),\beta_j) & \text{if } i = j \wedge \rec[\pi_j]{m} \notin A_j^\In \\
	   (s_j,\beta_j.m) & \text{if } \rec[\pi_j]{m} \in A_j^\In \wedge i \neq j\\
	   (\step_j(s_j,\snd{m}),\beta_j.m) & \text{if } i = j \wedge \rec[\pi_j]{m} \in A_j^\In \\
	    (s_j,\beta_j) & \text{otherwise}
	 \end{cases}
       \]
     \item if for a process $\pi_i$, there exists $\rec[\pi_i]{m} \in A_i^{\In}$ such that 
       \[\step_i(s_i,\rec[\pi_i]{m}) \neq \mathit{undef} \wedge \beta_i = \alpha_1.m.\alpha_2 \wedge \forall b \in \alpha_1\ \step_i(s_i,\rec[\pi_i]{b}) = \mathit{undef} \]
       then 
       \[\step( ( q_1,  \ldots, (s_i,\beta_i), \ldots,q_n), \rec[\pi_i]{m}) =
       ( q_1, \ldots, (\step_i(s_i,\rec[\pi_i]{m}),\alpha_1.\alpha_2), \ldots, q_n ) \]
   \end{itemize}
\item $D = \{\pi_1,\ldots,\pi_n\}$,
\item $\dom : A \rightarrow D$ such that $\dom(\snd{m})=\dom(\rec[\pi_j]{m})= \pi_i$ where $\pi_i$ is the only process such that $\snd{m} \in A^{out}_i$,
\item $\obs:D \times S \rightarrow \bigcup\limits_{i=1}^n (S_i \times A^\In_i)$, defined by $\obs(\pi_i, (q_1, \ldots, q_i, \ldots , q_n))=q_i$.
 \end{itemize}
\end{definition}

As for processes, a distributed machine cannot execute any action from any state.
We denote by $Exec(\machine)$ the valid executions of a distributed machine.
For a sequence $\alpha \in Exec(\machine)$, we denote by $s^0.\alpha$ the global state reached after executing $\alpha$.

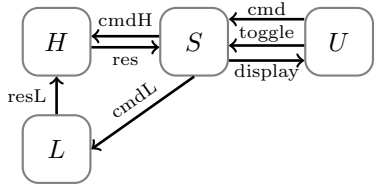
\begin{figure}[htpb]
  \centering
  \tikzstyle{process}=[minimum size=0.9cm,draw=gray,thick,rounded corners=6pt]
\begin{tikzpicture}
  \path[use as bounding box] (-0.2,0.2) rectangle (4.8,-2.4) ;
  \node[process,anchor=north west] (H) {$H$} ;
  \node[process,anchor=west] at ($(H.east) + (0.9,0)$)   (S) {$S$} ;
  \node[process,anchor=west] at ($(S.east) + (1.0,0)$)   (U) {$U$} ;
  \node[process,anchor=north] at ($(H.south) - (0,0.5)$) (L) {$L$} ;
  \def\excent{9}

  \node[coordinate] at (S.180-\excent) (cmdH) {} ;
  \node[coordinate] at (S.180+\excent) (dispH) {} ;
  \begin{scope}[line width=1pt,font=\scriptsize]
    \draw[<-] (cmdH -| H.east) node[coordinate] {} -- node[above] {cmdH}  (cmdH);
    \draw[->] (dispH -| H.east) node[coordinate] {} -- node[below] {res} (dispH);

    \draw[<-] ($(S.east) + (0,0.3)$) node[coordinate](n) {} -- node[above,inner sep=1pt] {cmd}  ($(U.west |- n) $);
    \draw[<-] ($(S.east) + (0,-0.05)$) node[coordinate](n) {} -- node[above,inner sep=1pt] {toggle}  ($(U.west |- n) $);
    \draw[->] ($(S.east) + (0,-0.25)$) node[coordinate] (n) {} -- node[below,inner sep=1pt] {display} (U.west |- n);

    \node[coordinate] at (L.north) (dispL) {} ;
    \draw[<-] (H.south)  -- node[left] {resL} (dispL) ;

    \node[coordinate] at (S.south) (scmdL) {} ;
    \node[coordinate] at (L.east) (rcmdL) {} ;
    \draw[->] (scmdL.north) -- node[sloped,above] {cmdL} (rcmdL.south) ;
  \end{scope}
\end{tikzpicture}
  \caption{A distributed machine obtained by composing the processes $\{H,L,S,U\}$. Each arrow corresponds to a message that can be exchanged between two processes.}
  \label{fig:machine}
\end{figure}

\begin{example}
  Figure~\ref{fig:machine} represents a global view of the Starlight Interactive Link. 
  There are four processes: a high-security network $H$, a low-security network $L$, the switch $S$ detailed in Figure~\ref{fig:process} and a user $U$.
  The user toggles between high-security and low-security mode by sending a toggle message.
  The user inputs commands (cmd) to the switch, which forwards them to the high- or low-security network, depending on the current mode. 
  Upon reception of a command, the $H$ or $L$ process executes it and outputs  the result (resL and res messages respectively).
  The result of a command executed in the low-security network is transmitted back to the switch through the high-security network.
  Upon reception of a result, the switch forwards it to the user through a display message.
  
  We do not detail the behavior of each component.
  The beginning of a valid execution of the system is $\snd{\text{cmd}}\ \snd{\text{toggle}}\ \rec[S]{\text{cmd}}\ \snd{\text{cmdH}}\ \rec[S]{\text{toggle}}$.
  Actions $\rec[S]{\text{cmd}}$, $\rec[S]{\text{toggle}}$ and $\snd{\text{cmdH}}$ correspond to the transitions of process $S$ as depicted in Figure~\ref{fig:process}.

  Each process is a security domain. 
  Each action that is sending or receiving a message is associated to the domain corresponding to the sender process.
  For instance,  $\dom(\snd{\text{cmdL}}) = \dom(\rec[L]{\text{cmdL}}) = S$. 
  Each process can observe its local state and its input buffer.
  For instance, in the global state $s.\snd{\text{cmd}} \snd{\text{toggle}}$ reached after executing $\snd{\text{cmd}} \snd{\text{toggle}}$, we have $\obs(S,s) = (h,\text{cmd } \text{toggle})$.
  
  In this example, the security property to ensure is ``The low security network has no information about the commands sent to the high security network''.
\end{example}

\subsection{Security policy}

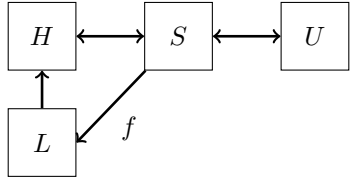
\begin{figure}[htbp]
  \centering
  \tikzstyle{process}=[minimum size=0.9cm,draw]
\begin{tikzpicture}
  \path[use as bounding box] (0,0) rectangle (4.6,-2.4) ;
  \node[process,anchor=north west] (H) {$H$} ;
  \node[process,anchor=west] at ($(H.east) + (0.9,0)$)   (S) {$S$} ;
  \node[process,anchor=west] at ($(S.east) + (0.9,0)$)   (U) {$U$} ;
  \node[process,anchor=north] at ($(H.south) - (0,0.5)$) (L) {$L$} ;

  \begin{scope}[line width=1pt]
    \draw[<->] (H) -- (S);
    \draw[<->] (S) -- (U);
    \draw[->] (S) -- node[anchor=north west] {$f$} (L.east) ;
    \draw[->] (L) -- (H) ;
  \end{scope}
\end{tikzpicture}
  \caption{Security policy with the domains of the machine from Figure~\ref{fig:machine}.}
  \label{fig:security_policy}
\end{figure}

A security policy is represented as a directed graph, whose vertices are the security domains and edges may be labeled by filter functions.
Intuitively, an edge between two security domains allows information to flow according to the direction of the edge. 
An example of security policy is depicted in Figure~\ref{fig:security_policy} which presents how information flows in the machine depicted in Figure~\ref{fig:machine}.

In our model, transmission of information is expressed through actions, that modify the state of different processes.
The presence of an edge between two security domains indicates that any action whose domain is the source of the edge may transmit information to the destination of the edge.
According to Figure~\ref{fig:security_policy}, an action executed in process $S$ might affect both the state of $H$, $L$ and $U$. 
An edge labeled by a filter function may allow or not an action at the source to affect the state of the destination, depending on the sequence of actions executed so far.
In the figure, the edge between $S$ and $L$ is labeled by a filter function $f$.
The intuitive role of $f$ is to ensure that $L$ receives only commands sent in low security mode by the user.

\begin{definition}
  Given a set of actions $A$, a filter function $f:A^* \rightarrow \{True,False\}$ returns a boolean indicating whether the last action of the sequence can be transmitted.

  We denote by $\filters(A)$ the set of filter functions defined on the set of actions~$A$.
\end{definition}

In our example, the function $f$ is true only when both:
\begin{itemize}
  \item the last action is $\snd{\text{cmdL}}$, and
  \item the number of toggle messages received is odd, indicating that the user is inputting commands in low security mode.
  (We assume that the system starts in high security mode.)
\end{itemize}
Formally, we write  $f(\alpha a) = (a=\snd{\text{cmdL}}) \wedge (|\alpha|_{\rec{toggle}} \mod 2 =1)$, where $|\alpha|_{a}$ is the number of occurrences of $a$ in the sequence $\alpha$.

\begin{definition}
  A \emph{security policy} $\secpol$ defined over the set of actions $A$ is a pair $(D,\leadsto)$ where 
  \begin{itemize}
    \item $D$ is a set of security domains and 
    \item $\leadsto \subseteq D \times \filters(A) \cup \{\top\} \times D$ is a set of edges, labeled by filter functions in $\filters$ or $\top$. 
      We require that:
      \begin{itemize}
	\item $\forall \pi_i \in D. (\pi_i, \top,\pi_i) \in \leadsto$.
	\item if $(\pi_i,f,\pi_j) \in \leadsto$, then $f$ is in $\filters(A^{\In}_i \cup A^{\out}_i)$.
      \end{itemize}
  \end{itemize}
\end{definition}

The first restriction on the definition of $\leadsto$ states that each security domain can always observe actions that are associated to itself.
The second restriction states that a filter function applied to the flow between a process $\pi_i$ and $\pi_j$ should depend only on the sequence of messages received and sent by $\pi_i$. 
It can be extended to a function depending on all actions executed on the system by simply discarding actions that are not in $A^\In_i \cup A^\out_i$.

We write $\pi_i \mathrel{\substack{f \\ \leadsto \\ ~}} \pi_j$ if $(\pi_i,f,\pi_j) \in \leadsto$.
An edge labeled by $\top$ imposes no restriction on the corresponding information flow.
In that case, we denote $\pi_i \leadsto \pi_j$.
A security policy is \emph{transitive} if it contains only edges labeled by $\top$ and the relation $\leadsto$ is transitive.
Otherwise, it is intransitive. 
We focus on intransitive policies, such as the one depicted in Figure~\ref{fig:security_policy}.

There are several constructions for representing the maximal information allowed by a security policy~\cite{chongmeyden12}, yielding different notions of security.
Contrarily to the model described in~\cite{chongmeyden12}, we do not assume that any action is possible from any state.
Consequently, each process knows the set of possible global executions and may infer some informations about the global state, based on its observation.
In our example, when the low security network receives a command to execute, it knows that this command was sent before by the user to the switch.
Such information may be computed by using knowledge with perfect recall as in~\cite{balliu13}.

The information available to a security domain after executing a given sequence of actions is obtained by purging actions not visible by the security domain.
An information is not visible by a security domain if either there is no incoming arrow from the domain of the action or if the incoming arrow is labeled by a filter function that evaluates to false.
Formally, we recursively define the purge function for a security domain $\pi$ as $\purge_\pi(\epsilon)=\epsilon$ and
\[ \purge_\pi(\alpha a) = 
  \begin{cases}
    \purge_\pi(\alpha) & \text{if } \dom(a) \not\leadsto \pi \vee (\dom(a) \mathrel{\substack{f \\ \leadsto \\ ~}} \pi \wedge \neg f(\alpha a) )\\
    \purge_\pi(\alpha)a &  \text{otherwise}
  \end{cases}
\]

The purged execution sequence represents the maximal information that a process is allowed to have at a given execution point.
A distributed machine complies with a security policy if for each security domain, the observation after executing a sequence $\alpha$ depends only on the purged sequence.



\begin{definition}
  \label{def:fta_sec}
  A distributed machine $\machine =(S,s^0,A,\step,D,\dom,\obs)$ complies with the security policy $\secpol=(D,\leadsto)$ if:
  \begin{align*}
    \forall \pi_i \in D,\forall \alpha, \beta \in Exec(\machine), & \\
    \purge_{\pi_i}(\alpha) = \purge_{\pi_i}(\beta)& \implies \obs_{\pi_i}(s^0.\alpha) = \obs_{\pi_i}(s^0.\beta)
  \end{align*}
\end{definition}

Finally, we remark that a distributed machine implicitly defines a security policy without filter functions.
The security domains are already defined by assigning a domain to each process.
If a message can be sent from a process $\pi_1$ to a process $\pi_2$, there is an arrow from the security domain $\pi_1$ to the security domain $\pi_2$.

\begin{definition}
  \label{def:implicit_secpol}
  The \emph{implicit security policy} defined by a distributed machine $\machine$ obtained by composing the processes $\{\pi_1,\ldots,\pi_n\}$ is the pair $(D,\leadsto)$ where:
   \begin{itemize}
     \item $D = \{\pi_1,\ldots,\pi_n\}$ each process is a security domain,
     \item $\snd{m} \in A_i^{\out} \wedge \rec[\pi_j]{m} \in  A_j^\In \implies (\pi_i,\top,\pi_j) \in \leadsto$. 
       If a message $m$ can be sent from $\pi_i$ to $\pi_j$, there is an arrow between the corresponding security domains.
   \end{itemize}
\end{definition}

\section{Proving Security Policies for Distributed Machines}
\label{sec:unwinding}

We use the method based on unwinding relations~\cite{gm84} to prove that a machine complies with a given security policy.
An unwinding relation assigns to each security domain $\pi$ an equivalence relation on the states of the machine that we denote by $\sim_\pi$.

We first consider the case where all edges of the security policy are labeled by $\top$, that is without filter functions.
In that case, the Theorem~1 by Rushby~\cite{rushby1992} states that a machine $\machine$ complies with a security policy if there exists a family of unwinding relations $\{\sim_\pi\}_{\pi \in D}$ such that, for any two states $s,t \in S$, any process $\pi \in D$ and any action $a \in A$, we have:
\begin{itemize}
  \item \emph{Output consistency.}   
    If $s \sim_\pi t$, then  $ \obs(\pi,s) = \obs(\pi,t)$.
  \item \emph{Step consistency.} If $ s \sim_\pi t$ then $\forall a \in A,
    \step(s,a) \sim_\pi \step(t,a)$.
  \item \emph{Local respect.} If $\mathtt{dom}(a) \not\leadsto \pi$, then $s \sim_\pi \step(s,a)$.
\end{itemize}
Note that we actually use the formulation by van der Meyden~\cite{meyden07}.
Using unwinding relations, we prove that a distributed machine complies with its implicit security policy described in Definition~\ref{def:implicit_secpol}.

\begin{proposition}
  \label{prop:implicit_comply}
  A distributed machine complies with its implicit security policy.
\end{proposition}

The proof relies on the fact that the current state of a process depends only on the messages it receives and sends. 
The implicit security policy allows each process to observe these actions.
Therefore the distributed machine complies with the implicit security policy.

\proof We consider the unwinding relations obtained as follows:
Given two states $q=(q_1,\ldots,q_n)$ and $r=(r_1,\ldots,r_n)$, we define $q \sim_{\pi_i} r$ iff $ q_i = r_i$.
This is clearly an equivalence relation.
Recall that for each $i$, $q_i = (s_i,\beta_i)$ where $s_i$ is the state of the process and $\beta_i$ the input buffer.
We now prove the three properties needed to establish security:
\begin{itemize}
  \item \emph{Output consistency} $q \sim_\pi r \implies q_i = r_i \implies \obs(\pi_i,q) = \obs(\pi_i,r)$.
  \item \emph{Step consistency} Let $a \in A$ be an action, $q,r \in S$ be two global states and $q'=\step(q,a),$ $r'=\step(r,a)$ the states reached when executing $a$. We assume that $q\sim_\pi r$, that is $q_i = r_i$. Recall that $a$ is either the emission of a message $\snd{m}$ or the reception of a message $\rec[\pi_i]{m}$.

    If $\pi_i$ is not sender nor receiver of the message $m$, by definition of $\step$, we have $q'_i = q_i$ and $r'_i = r_i$. 
    Since $q_i = r_i$, we have  $q' \sim_{\pi_i} r'$.
    
    If $a=\snd{m}$ is the emission of the message $m$:
    \begin{itemize}
      \item if $\pi_i$ is the sender of the message,
	by definition of $\step$, $q'_i = (\step_i(s_i,m),\beta_i)$ and $r'_i = (\step_i(t_i,m),\gamma_i)$. 
        Since $(s_i,\beta_i) = q_i = r_i = (t_i,\gamma_i)$, we have $q'_i = r'_i$ that is $q' \sim_{\pi_i} r'$.
      \item if $\pi_i$ is a receiver of the message,
	by definition of $\step$, $q'_i = (s_i,\beta_i m)$ and $r'_i = (t_i,\gamma_i m)$. 
        Since $(s_i,\beta_i) = q_i = r_i = (t_i,\gamma_i)$, we have $q'_i = r'_i$ that is $q' \sim_{\pi_i} r'$.
    \end{itemize}
    If $a=\rec[\pi_j]{m}$ is the reception of the message $m$:
    \begin{itemize}
      \item if $\pi_i$ is the sender of the message,
	by definition of $\step$, $q'_i = q_i$ and $r'_i = r_i$, that is $q' \sim_{\pi_i} r'$.
      \item if $\pi_i$ is the receiver of the message,
	by definition of $\step$, 
	$q'_i = (\step_i(s_i),$ $\alpha_1\alpha_2)$ and $r'_i = (\step_i(t_i),\alpha'_1\alpha'_2)$, where $(s_i,\alpha_1 m \alpha_2) = q_i = r_i = (t_i,\alpha'_1 m \alpha'_2)$.
	Since $m$ does not appear in $\alpha_1$ or $\alpha'_1$, we have $q'_i = r'_i$ that is $q' \sim_{\pi_i} r'$.
    \end{itemize}
  \item \emph{Local respect} 
    Let $a \in A$ be an action, $q \in S$ be a state, and $q'=\step(q,a)$.
    By definition of $\leadsto$, $\dom(a) \not\leadsto \pi_i$ only if $\dom(a)$ does not send any message to $\pi$.  
    By definition of $\step$, if $a$ is the emission or the reception of a message not involving $\pi_i$, then $q'_i = q_i$.
    Thus $q \sim_{\pi_i} \step(q,a)$.  \qed
\end{itemize}



We extend the unwinding theorem to security policies with filter functions.
We define an additional property, that depends only on the local state of the component.
This property ensures that the transmission of a message from a process $\pi_i$ to a process $\pi_j$ cannot take place whenever the filter function on the edge between security domains $\pi_i$ and $\pi_j$ evaluates to false. 
The transmission of the message cannot take place if either $\pi_j$ is not a receiver of the message or the send action is not possible in $\pi_i$.

\begin{itemize}
  \item \emph{Local Filter Respect}: 
    $\forall (\pi_i,f,\pi_j) \in \leadsto\forall \delta \in LocalExec(\pi_i)$ \\
        $\neg f(\delta a) \implies a \notin A_j^\In \vee \step(s^0_i.\delta,a) = \mathit{undef}$
\end{itemize}

Theorem~\ref{thm:unwinding_filter} states that a distributed machine complies with a security policy provided that there exists an unwinding relation that respects output consistency, step consistency, local respect and that local filter respect is also verified.

\begin{theorem}
  \label{thm:unwinding_filter}
  Let $\machine$ be a machine, $\secpol$ be a security policy, and for each domain $\pi$ $\sim_\pi$ be an equivalence relation.
  If the relations $\{\sim_\pi\}_{\pi \in D}$ verify output consistency, step consistency, local respect and local filter respect properties, then the machine  $\machine$ complies with the architecture $\secpol$.
\end{theorem}
\proof Let $\alpha,\beta \in Exec(\machine)$ be two executions of $\machine$.
We prove by induction on $|\alpha| + |\beta|$ that $\purge_\pi(\alpha) = \purge_\pi(\beta) \implies s^0.\alpha \sim_\pi s^0.\beta$.

The base case is $\beta = \alpha = \epsilon$, and we have $s^0 \sim_\pi s^0$.

Let us write $\alpha = \alpha' a$. We distinguish between the two following cases:
\begin{itemize}
  \item $\dom(a) \not\leadsto \pi$ or 
  $\dom(a) \mathrel{\substack{f \\ \leadsto \\ ~}} \pi$ and $f(\alpha)$ is false.
    In that case,  $\purge_\pi(\alpha a) = \purge_\pi(\alpha) = \purge_\pi(\beta)$. 
    By applying the induction hypothesis on $\alpha'$ and $\beta$, we obtain $ s^0.\alpha' \sim_\pi s^0.\beta$.
    Since $\alpha' a$ is a valid execution sequence, the local respect ensures that $a$ is not an action that can be received by $\pi$.
    Thus $s^0.\alpha' \sim_\pi \step(s^0.\alpha',a)$ and we conclude $s^0.\alpha \sim_\pi s^0.\beta$.
  \item $\dom(a) \leadsto \pi$ or 
  $\dom(a) \mathrel{\substack{f \\ \leadsto \\ ~}} \pi$ and $f(\alpha)$ is true.
  In that case, we can assume that $\beta$ also ends with $a$. 
  Otherwise, by swapping $\alpha$ and $\beta$, one falls back in the previous case.
  We write $\beta= \beta'a$.
  By definition of the purge function we have $\purge_\pi(\alpha) = \purge_\pi(\alpha')a$ and similarly $\purge_\pi(\beta) = \purge_\pi (\beta')a$ and therefore $\purge_\pi(\alpha') = \purge_\pi(\beta')$.
  The induction hypothesis applied to $\alpha'$ and $\beta'$ gives us $s^0.\alpha' \sim_\pi s^0.\beta'$.
  The step consistency allows us to conclude $s^0.\alpha \sim_\pi s^0.\beta$.
\end{itemize}

We proved that $\purge_\pi(\alpha) = \purge_\pi(\beta) \implies s^0.\alpha \sim_\pi s^0.\beta$.
By using the output consistency, we have $s^0.\alpha \sim_\pi s^0.\beta \implies \obs_\pi(s^0.\alpha) = \obs_\pi(s^0.\beta)$, which concludes the proof. \qed

We already exhibited unwinding relations for the particular case of distributed systems, that prove compliance of a distributed machine with its implicit security policy.
In particular, a security policy obtained by labeling some edges of the implicit security policy with filter functions can be ensure by checking the local filter respect.
As the latter property involves only the actions of one process, it can be checked locally on that process.

\begin{example}
  Consider again the Starlight example.
  The security policy depicted in Figure~\ref{fig:security_policy} is obtained from the implicit security policy of the starlight example, with a additional filter function on the edge between $S$ and $L$.
  It is sufficient to check that $S$ respects the filter function $f$ to ensure that the system complies with the security policy from Figure~\ref{fig:security_policy}.
\end{example}

\section{Case Study}
\label{sec:case_study}
\looseness=-1

The case study discussed in this paper is a simplified version of the smart microgrid system that is implemented in our research lab \cite{koss2012}. The system contains a smart micro grid ($\mathit{SMG}$) that coordinates a finite set of prosumers $\mathit{Pr_1,\ldots,Pr_n}$. Each prosumer can produce energy and consume energy from the local production and from the grid. Moreover, each prosumer has the possibility to store energy in batteries. Therefore, a prosumer may sell energy or buy energy from the grid.

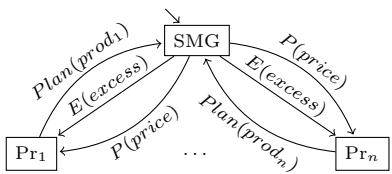
\begin{figure}[!ht]
\begin{center}
  \begin{tikzpicture}
  \def\ofs{1.3}
  \begin{scope}[font=\scriptsize]
  \node[draw,initialUp] (r) {SMG};
  \node[draw] at ($(r.south) + (-2.2,-\ofs)$) (b) {Pr$_1$}
    edge[pre,bend right] node [sloped,below,inner sep=1pt] {$P(price)$} (r) 
    edge[pre] node [sloped,above,inner sep=1pt] {$E(excess)$} (r) 
    edge[post,bend left] node [sloped,above,inner sep=1pt] {$Plan(prod_1)$} (r);
  \node[ellipse] at ($(r.south) + (0,-\ofs)$) (d) {\ldots};
  \node[draw] at ($(r.south) + (2.2,-\ofs)$) (c) {Pr$_n$}
    edge[pre,bend right] node [sloped,above,inner sep=1pt] {$P(price)$} (r) 
    edge[pre] node [sloped,above,inner sep=1pt] {$E(excess)$} (r) 
    edge[post,bend left] node [sloped,below,inner sep=1pt] {$Plan(prod_n)$} (r) ;
    \end{scope}
\end{tikzpicture}
  \caption{A machine of the smart microgrid system}
  \label{fig:pros_machine}
\end{center}
\end{figure}

Figure \ref{fig:pros_machine} shows a high-level view of the machine of our case study. The $\mathit{SMG}$ generates a price for energy and transfers it to the prosumers through action $\textit{P(price)}$. The price is the same for both buying and selling energy to the grid. Based on the price received and an estimation of the local production and consumption, each prosumer computes a production plan (variable $\mathit{prod_i}$ in the figure). This plan specifies the amount of energy the prosumer buys or sells from/to the grid. 
If this value is positive, then the prosumer produces more energy than its local needs which is sold to the grid, otherwise, it buys energy from the grid. The production plans are sent through the $Plan$ action to the $\mathit{SMG}$. When all prosumers have sent their production plans, the $\mathit{SMG}$ computes the global production of the grid. The global production may be negative, in which case the prosumers consume more energy than what they globally produce. 

The stability of the grid is assured if the global production does not exceed the available line capacity that is specified by an upper and a lower bound i.e. $\mathit{U_B}$, and $\mathit{L_B}$ respectively. Variable $\mathit{excess}$ returns the amount by which the global production exceeds the bounds of the line capacity. If the global production is within the bounds, the grid will be stable and the $\mathit{excess}$ variable has the value zero. Otherwise, it returns the amount by which the global production exceeds the bounds. 
If action $\mathit{E(excess)}$ transfers a nonzero value, all prosumers have to adjust their plans. After that, new plans are sent back to the $\mathit{SMG}$, which updates the excess and sends it to the prosumers. It may take more rounds to have the plans accepted by the $\mathit{SMG}$. Once the plans are validated, we assume that they become active for a finite period of time, after which a new price is generated and new plans are computed.

\begin{figure}[!ht]
\begin{center}
  \begin{tikzpicture}
  \def\ofs{1.3}
  \begin{scope}[font=\scriptsize]
  \node[draw,initialUp] (r) {SMG};
  \node[draw] at ($(r.south) + (-2.2,-\ofs)$) (b) {Pr$_1$}
    edge[pre,bend right] node [sloped,below,inner sep=1pt] {$f_{excess}$} (r) 
    edge[post,bend left] node [sloped,above,inner sep=1pt] {} (r);
  \node[ellipse] at ($(r.south) + (0,-\ofs)$) (d) {\ldots};
  \node[draw] at ($(r.south) + (2.2,-\ofs)$) (c) {Pr$_n$}
    edge[pre,bend right] node [sloped,above,inner sep=1pt] {$f_{excess}$} (r) 
    edge[post,bend left] node [sloped,below,inner sep=1pt] {} (r) ;
    \end{scope}
\end{tikzpicture}
  \caption{Security policy of the system}
  \label{fig:architecture}
\end{center}
\end{figure}
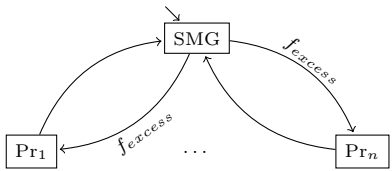

The security property that the smart microgrid system should ensure is that none of the prosumers can guess the plan of any other prosumer. 
Otherwise, a prosumer could change its own plan according to other prosumer plans and earn more profit and energy from the grid.
Figure \ref{fig:architecture} displays the security policy that expresses this security property. Since $\mathit{SMG}$ is the only trusted process, we specify  a filter function $\mathit{f_{excess}}$ on the edges from the $\mathit{SMG}$ to each of the prosumers.

Two auxiliary functions are needed for formally describing this filter function.
 First, given a finite sequence of actions $\mathit{\alpha=a_1,\ldots,a_m}$, for $\mathit{n\leq m}$, let $\mathit{suffix(\alpha,n)}$ 
denote the suffix of size n of $\alpha$, which is compound of the last $n$ actions of $\alpha$:
\begin{eqnarray*}
\mathit{suffix(\alpha,n)=a_{m-n+1},\ldots,a_m}.
\end{eqnarray*}
Second, given a finite sequence of actions $\alpha$ and an action $\mathit{a}$, the function $|\alpha |_\mathit{a}$ returns the number of occurrences of action $\mathit{a}$ in $\alpha$\@.
The global production $\mathit{Prod}$  is just the sum of the production plans of the individual prosumers: 
\begin{align*}
\mathit{Prod = \sum\limits_{i=1}^n prod_i}
\end{align*}
Using these auxiliary functions, the filter function $f_{excess}$ is defined as follows:
\begin{eqnarray}
\mathit{f_{excess}(\alpha a)} & = &\nonumber \\ 
	& &  \biggl(\mathit{a = P(price)} \wedge (\alpha=\epsilon \vee \mathit{suffix}(\alpha, 1)=E(0))\biggr) \vee \\
	& &   \biggl( \mathit{a=E(excess)} \wedge \bigwedge\limits_{\mathit{i}=1}^\mathit{n} |\mathit{suffix}(\alpha, \mathit{i})|_{\mathit{Plan(prod_i)}=1} \nonumber \\
	&& \wedge~\mathit{excess} = \mathit{compute\_excess(Prod)}\biggr)
\end{eqnarray}
Part (1) states that any $P$ action is either the initial action of the $\mathit{SMG}$ or it follows the emission of the
message indicating that the production plans do not exceed the line capacity bounds. 
Part (2) requires that whenever action $\mathit{E}$ occurs, each prosumer has sent its production plan exactly once in 
the last $\mathit{n}$ actions of the sequence $\alpha$ and the filter function sends the correct excess value $\mathit{compute\_excess(Prod)}$, defined as:
\begin{align*}
\mathit{compute\_excess(Prod)}= 
\begin{cases} 
  0,                                                & \text{if}~\mathit{L_B \leq Prod \leq U_B},\\
\mathit{Prod - U_B}, & \text{if}~\mathit{U_B <  Prod},\\
\mathit{Prod - L_B}, & \text{otherwise}.
\end{cases}
\end{align*}
The first condition holds whenever the global production does not exceed the line capacity bounds $\mathit{L_B}$ and $\mathit{U_B}$\@. 
The second case describes the case when prosumers produce too much energy,  and the third condition holds whenever 
the global production exceeds the line capacity lower bound, in which case too much energy is consumed.

The filter function $\mathit{f_{excess}}$ ensures the security property that no prosumer can deduce the production plan of any other prosumer.
Indeed, a prosumer only obtains one excess value after emitting one plan. 
This value depends on $\sum_{\mathit{i}=1}^n \mathit{prod_i}$ and therefore, the prosumer $\mathit{Pr_i}$ can obtain the value $\sum_{\mathit{j\neq i}} \mathit{prod_j}$, provided it knows the bounds $\mathit{L_B}$ and $\mathit{U_B}$.
However, for $\mathit{n} \neq 2$, the prosumer cannot deduce any information about the particular consumption of another given prosumer, since the value of $\sum_{\mathit{i\neq j}} \mathit{prod_i}$ could have be obtained from any other values $\mathit{prod'_i}$ such that $\sum_{\mathit{i\neq j}} \mathit{prod'_i} = \sum_{\mathit{i\neq j}} \mathit{prod_i}$. 

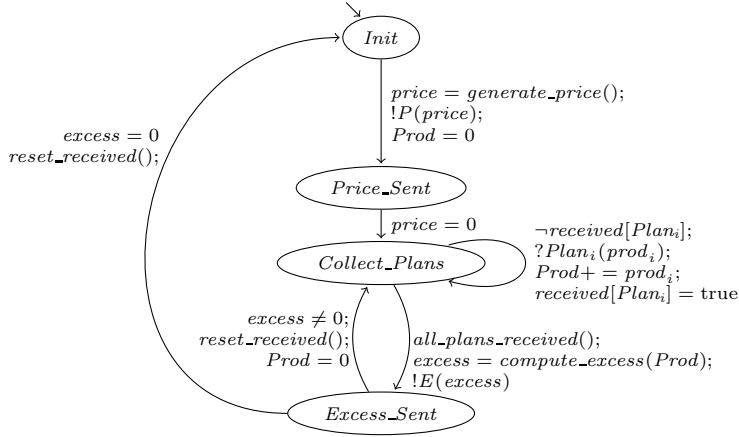
\begin{figure}[!ht]
\begin{center}
  \tikzstyle{place}+=[minimum size=4.5mm]
\begin{tikzpicture}
  \def\xsize{0.3}
  \def\ysize{1}
  \def\posl{0.6}
  \def\ofs{1.3}
  \def\margtop{0.25}

  \begin{scope}[font=\scriptsize]
    \node[draw, ellipse,initialUp] (a) {$\mathit{Init}$} ;
    \node[draw, ellipse] (d) at ($(a) - 2*(0,\ysize)$) {$\mathit{Price\_Sent}$} 
    edge[pre] node [right,align=left] (lftR) {$\mathit{price}=\mathit{generate\_price}()$;\\!$\mathit{P(price)}$;\\$\mathit{Prod}=0$} (a); 
    \node[draw, ellipse] (b) at ($(d) - 2*(0,0.5)$) {$\mathit{Collect\_Plans}$} 
    edge[pre] node [right,align=left] (lftR) {$\mathit{price}=0$} (d) 
      edge[pre,loop right, align=left] node   [right] (rht) {$\neg \mathit{received[Plan_i]}$;\\$?\mathit{Plan}_i(\mathit{prod}_i)$;\\$\mathit{Prod}+=\mathit{prod}_i$;\\$\mathit{received[Plan_i]}=\text{true}$} (b) ;
      \node[draw, ellipse] (c) at ($(b) - 2*(0,\ysize)$) {$\mathit{Excess\_Sent}$}
    edge[pre,bend right,align=left] node [below right,pos=\posl,inner sep=1pt] (rhtR) {$\mathit{all\_plans\_received}$();\\$\mathit{excess=compute\_excess(Prod)}$;\\
    $\mathit{!E(excess)}$\\} (b)
    edge[post,bend left,align=left] node [align=right,left,inner sep=1pt] {$\mathit{excess}\neq 0$;\\$\mathit{reset\_received}()$;\\$\mathit{Prod}=0$} (b)
    edge[post,out=180,in=180,looseness=1.5] node [above left,pos=\posl,inner sep=1pt,align=right] {$\mathit{excess}=0$\\$\mathit{reset\_received}()$;} (a);
    \coordinate (tlR) at ($(lftR.west |- a.north) + (-0.1,\margtop)$) ;
    \coordinate (brR) at ($(rht.east |- c.south) + (0.1,-0.1)$) ;
  \end{scope}
\end{tikzpicture}
  \caption{Implementation of the $\mathit{SMG}$}
  \label{fig:uppaal_model}
\end{center}
\end{figure}

According to Theorem~\ref{thm:unwinding_filter} and Proposition~\ref{prop:implicit_comply}
it is sufficient to check that the process $\mathit{SMG}$ respects the filter function $\mathit{f_{excess}}$ for proving that
the overall systems enforces the security policy of Figure \ref{fig:architecture}\@. 

We are using model checking for automatically discharging this verification condition
based on an Uppaal model \cite{DBLP:conf/qest/BehrmannDLHPYH06} of the smart microgrid
and CTL encodings of the local verification condition for process $\mathit{SMG}$.
Figure \ref{fig:uppaal_model} displays the automaton of the $\mathit{SMG}$ process. 
The local states of this automaton are pairs of locations and valuations of variables. 
Deadlines for the time the $\mathit{SMG}$ waits for prosumer plans to arrive could be added
to the Uppaal model in a straightforward way. 

The $\mathit{SMG}$ automaton in Figure \ref{fig:uppaal_model} has
 $\mathit{Init}$ as the initial location and a transition on which the energy price is generated and sent to prosumers, 
$\mathit{Price\_Sent}$ is reached after the price is sent to prosumers and after which we reset the value of the energy price, 
and $\mathit{Collect\_Plans}$ is active for the period when plans are collected. 
Channel $\mathit{Plan_i}$ is binary, meaning that each prosumer sends its production plan on a unique channel. 
Guard $\mathit{\neg received[Plan_i]}$ guarantees that the $\mathit{SMG}$ takes into account the first production plan each prosumer sends. 
Here received is an array of booleans that records which prosumers have already sent their plans. 
When a prosumer tries to send a new plan before receiving an $\mathit{E}$ action, the $\mathit{SMG}$ simply ignores this action. 
Local variable $\mathit{Prod}$ adds up the plans received. 
When the $\mathit{SMG}$ has received a plan from each prosumer (i.e. function $\mathit{all\_plans\_received}$ returns true), the transition 
to location $\mathit{Excess\_Sent}$ is taken. 
On this transition, the excess is computed using function $\mathit{compute\_excess}$ and sent to prosumers. 
When the excess is not zero, the transition between $\mathit{Excess\_Sent}$ and $\mathit{Collect\_Plans}$ is taken which makes the 
$\mathit{SMG}$ ready to receive adjusted production plans. 
Finally, in case the excess is zero, the transition to $\mathit{Init}$ is taken. 
In both cases, the array of boolean indicating which plans have been received is reset through the $\mathit{reset\_received}()$ function.

In order to check that the $\mathit{SMG}$ respects the filter function, one simply needs to check:
 \begin{align*}
\forall \alpha \in \mathit{LocalExec(SMG)}, &\forall \mathit{Pr_i} \in \{\mathit{Pr_1,\ldots,Pr_n}\} :\\
&\neg \mathit{f(\alpha a)} \implies \mathit{a} \notin \mathit{A}^{\In}_{\mathit{Pr_i}} \vee \step_{\mathit{SMG}}(\mathit{s^0_{SMG}}.\alpha, \mathit{a}) = \mathit{undef}.
 \end{align*}
 That is, for every sequence of actions $\alpha$ and an action $a$ that falsifies the filter function, either the action $a$ is not transmitted to the prosumers or the action $a$ is not possible in the state reached after executing $\alpha$. By contraposition, we have to check that if the action $a$ is an input for the prosumer and it can be executed after $\alpha$, then the filter function evaluates to true. Formally:
  \begin{align*}
 a \in A^{\In}_{Pr_i} \wedge \step_{SMG}(s^0_{SMG}.\alpha, a) \neq \mathit{undef} \implies f(\alpha a)
  \end{align*}
 
In our case, the messages that can be sent to a prosumer are E(excess) and P(price).

\begin{itemize}
  \item We show that when action $\mathit{P(price)}$ is sent, then Part (1) of the filter function is satisfied.
When action $\mathit{P}$ is fired, either a) no other action has occurred in process $\mathit{SMG}$ or b) the production plans do not exceed the line capacity bounds. In the former case, all the variables are initialized to default values, meaning that excess is initialized to zero. The latter case implies the same, i.e. location $\mathit{Price\_Sent}$, from where action $\mathit{P(price)}$ is sent, is reached only when excess is zero. These allows us to write the following formula:

    \begin{align*}
      AG (\mathit{Price\_Sent} \implies (\mathit{excess}=0)),
    \end{align*}
 where $\mathit{Price\_Sen}t$ is the location reached immediately after action $\mathit{P(price)}$ is sent.
  \item We show that when $\mathit{E(excess)}$ is sent, then Part (2) of the filter function is satisfied.
   First, location $\mathit{EXCESS\_SENT}$ is reached whenever the excess action has been sent. Whenever the excess is computed, we can easily see in the model of $\mathit{SMG}$ that the $\mathit{Prod}$ variable counts the first plan received from each prosumer(guard $\mathit{\neg received[Plan_i]}$), but each prosumer has sent a plan (guard $\mathit{all\_plans\_received}$). Therefore the second part of the filter function, which requires that global production plan includes a single plan received from each prosumer, is encoded in the model and there is no need to formally verify it. It remains to reason about the value of the excess function transmitted to prosumers. 
The following property checks the value of the excess sent on the $\mathit{E}$ action:
    \begin{align*}
      AG (\mathit{EXCESS\_SENT} \implies (&(\mathit{excess}= 0 \wedge \mathit{Prod \leq U_B \wedge Prod\geq L_B}) \vee \\
  &   (\mathit{excess = Prod - U_B \wedge Prod > U_B}) \vee \\
  &  (\mathit{excess = Prod - L_B \wedge Prod < L_B}))).
    \end{align*}
\end{itemize}

The Uppaal model built\footnote{The model is available upon request.} is parametrized with the number of prosumers. Therefore, we have easily analyzed different configurations for prosumers. The verification is fast e.g. a configuration with 100 prosumers verified in less than 1 second. The Uppaal version used is 4.1.14 (64-bit), run on a Macbook Pro with 8GB of memory.  

In summary, we demonstrated how a security policy with filter function can be defined for ensuring the security of a smart microgrid system.  Moreover, we have shown a transformation of the verification condition for the trusted $\mathit{SMG}$ component and its associated 
filter function into a  CTL property,  which can readily be checked using standard model checking techniques. 

\section{Related Works}
\label{sec:related}

The notion of security policy, non-interference and the $\purge$ function goes back to the seminal work of Goguen and Meseguer~\cite{gm82}\@.
A purged sequence is a valid execution sequence for the global model, and non-interference establishes that a security domain is not able to 
distinguish between the execution of the original or the purged sequence.
In contrast, our purged sequences are not valid executions of the global system in that they contain only actions directly visible by a process,
and corresponding  security properties state that two security domains should not distinguish two executions that have the same purged sequence. 
Our developments also build on the notion of unwinding relations~\cite{gm84} and Rushby's unwinding theorem~\cite{rushby1992}, which
we generalize to security policies with filter functions.

An intransitive non-interference security policy assumes that all information available to a security domain is possibly transmitted 
whenever an action of the domain is executed (Van der Meyden~\cite{meyden07}, Rushby~\cite{rushby1992})\@. 
Chong and van der Meyden ~\cite{chongmeyden12} extend this framework to filter functions in order to limit the information 
transmitted about the history. This requires variations on the $\purge$ function depending on the notion of security used. 
Since a sequence of receive and send messages  fully determines the state of the distributed system, the original $\purge$ function 
is sufficient in our case.

Mantel~\cite{mantel00} and Balliu~\cite{balliu13} consider systems defined by a given set of execution traces.
In these systems, a given action is not always possible in contrast to Van der Meyden~\cite{meyden07} and Rushby's~\cite{rushby1992} setting\@.
Mantel's~\cite{mantel00} development work for transitiv security policies, whereas our work focuses on intransitive policies.

Balliu~\cite{balliu13} considers distributed systems for which the confidentiality of channels between processes needs to be ensured.
The condition for security is based on knowledge with perfect recall; low security components should not be able to infer the 
content of a high security channel based on the history of observation on the low security channels.
In the Starlight example, if the user always requests each command once in high security mode and once in low security mode, 
the low security network knows all executed high security commands. 
Here, we are considering such an implementation is to be secure, since the filter function is respected, 
but it is insecure in Balliu's~\cite{balliu13} framework\@. 
This requires, however, to exactly know the set of traces of the system, and thus the concrete implementation of every process.
In other words, such properties need to be checked on the global model, whereas we reduce the checking of global security policies
to verification conditions on single security domains.
It seems to be interesting to investigate which knowledge-based security properties can actually be encoded in 
terms of security policies.

Zhang~\cite{zhang11} describes filter functions in terms of regular expressions, and proves a result comparable
to  Theorem~\ref{thm:unwinding_filter}\@. Furthermore, intransitive security policies are 
encoded  through filter functions. Zhang~\cite{zhang11} framework, however, does not cover the case of  distributed 
systems consisting of  processes  communicating through messages, where each process constitutes a separate security domain.
Thus, checking that the system respects a filter function needs to be done globally, whereas we check each filter function against a single process.

\section{Conclusions and future work}
\label{sec:conclusion}

Our main results reduces compliance proofs of distributed system with intransitive security policies, 
possibly strengthened by filter functions, to localized verification conditions for each filter function.
These localized conditions only require proving that  source  processes filter function obey their filter functions. 
In this way, security proofs for distributed systems are decomposed into localized verification conditions. 
Our work is motivated by the development of a distributed separation kernel in the D-MILS project, which allows
us to deploy security policies on a  distributed computing platform.

We have applied our techniques to a non-trivial case study involving a prosumer-based smart microgrid, and applied CTL model 
checking for automatically checking the localized verification conditions. It would be interesting to enrich filter functions with 
further information about grid stability, optimization of energy consumption\cite{Katoen}, or timing constraints\cite{Dimitrova}\@. 
For the smart microgrid case study we have manually derived the CTL verification conditions, but it should be possible to automatically 
generate these temporal logic formulas from the definition of filter functions. 
It also seems to be worthwhile to investigate how filter functions may be extracted automatically from formally defined  security properties. 

Further work towards applying our framework to  the development and deployment of real-world security applications include the
extension of filter functions with cryptographic information, in particular, using the symbolic approach Dolev-Yao model \cite{dolevyao81}\@. 
Moreover, sequential dependencies between filter functions may also need to be considered. Because of the localized verification 
conditions for demonstrating security it is conceivable to  provide for incremental security proofs in  a setting with dynamically 
changing security policies and distributed systems.

\bibliographystyle{splncs03}
\bibliography{bib}

\end{document}